\documentclass[%
 reprint,
 amsmath,amssymb,
  aps,
 prb,
]{revtex4-1}

\usepackage{graphicx}
\usepackage{subfigure}
\usepackage{dcolumn}
\usepackage{bm}
\usepackage{hyperref}

\begin{document}

\preprint{APS/123-QED}

\title{Dynamical relaxation behavior of extended XY chain with gapless phase following a quantum quench}

\author{Kaiyuan Cao}%
 \affiliation{%
 Zhejiang Lab, Hangzhou 311100, P. R. China
}%

\author{Yayun Hu}
 \email{hyy@zhejianglab.com}
\affiliation{Zhejiang Lab, Hangzhou 311100, P. R. China}

\author{Peiqing Tong}
 \email{pqtong@njnu.edu.cn}
\affiliation{Department of Physics and Institute of Theoretical Physics, Nanjing Normal University, Nanjing 210023, P. R. China}
\affiliation{Jiangsu Key Laboratory for Numerical Simulation of Large Scale Complex Systems, Nanjing Normal University, Nanjing 210023, P. R. China}

\author{Guangwen Yang}%
 \email{ygw@tsinghua.edu.cn}
\affiliation{Zhejiang Lab, Hangzhou 311100, P. R. China}
\affiliation{Department of Computer Science and Technology, Tsinghua University, Haidian District, Beijing 100084, P. R. China}%

\date{\today}

\begin{abstract}
  We investigate the dynamical relaxation behavior of the two-point correlation in extended XY models with a gapless phase after quenches from various initial states. Specifically, we study the XY chain with gapless phase induced by the additional interactions: Dzyaloshinskii-Moriya interaction and XZY-YZX type of three-site interaction. When quenching from the gapped phase, we observe that the additional interactions have no effect on the relaxation behavior. The relaxation behavior is $\delta C_{mn}(t)\sim t^{-3/2}$ and $\sim t^{-1/2}$ for the quench to the commensurate phase and the incommensurate phase, respectively. However, when quenching from the gapless phase, we demonstrate that the scaling behavior of $\delta C_{mn}(t)$ is changed to $\sim t^{-1}$ for the quench to the commensurate phase, and the decay of $\delta C_{mn}(t)$ follows $\sim t^{-1}$ or $\sim t^{-1/2}$ for the quench to the incommensurate phase depending on the parameters of pre-quench Hamiltonian. We also establish the dynamical phase diagrams based on the dynamical relaxation behavior of $\delta C_{mn}(t)$ in the extended XY models.
\end{abstract}

\maketitle


\section{Introduction}

Advancements in ultra-cold atomic experiments have sparked significant interest in nonequilibrium many-body physics \cite{nature.2002.419.51, nature.2012.8.325, science.2012.337.1318, Oxford.2012, nature.2015.11.124}. One particularly important issue of this field is the investigation of the nonequilibrium time evolution of isolated quantum systems over long time scales \cite{AdvPhys.2010.59.1063, RevModPhys.83.863, PRB.2013.87.245107, AdvPhys.2016.65.239}. Numerous studies have focused on the dynamical relaxation of different physical quantities, such as the entanglement entropy \cite{PhysRevB.2016.94.214301, JPA2014.47.255301, JSM2016.064003, PhysRevE.2018.98.032110, PhysRevB.2023.107.134201}, two-point longitudinal correlation function \cite{JPA2014.47.175002, PRL2021.126.210602, PRB2022.105.094304}, population imbalance \cite{PRB2017.96.054303}, antiferromagnetic order parameter \cite{PRL2020.124.130602}, and ferromagnetic order parameter \cite{PRL2022.128.050601}. These studies collectively contribute to a comprehensive understanding of dynamical relaxation in a wide range of physical systems.

Recently, a class of dynamical phase transitions, characterized by the relaxation behavior of the two-point correlation $C_{mn}(t)=\langle\psi(t)|c_{m}^{\dag}c_{n}|\psi(t)\rangle$, has been proposed in periodically driven systems \cite{PhysRevB.2016.94.214301, JPA.2018.51.334002, PRB2020.102.235154, PhysRevB.2022.105.104303}. The difference $\delta C_{mn}(t)=C_{mn}(t)-C_{mn}(\infty)$ of the correlation at time $t$ from their steady-state values decays as a power law behavior $t^{-\mu}$. The transition of scaling exponent $\mu$ is claimed to characterize the dynamical phase \cite{PhysRevB.2016.94.214301}. Later, this type of relaxation behavior has also been observed in the systems after a quantum quench \cite{PRB.2022.105.054301, PhysRevB.2023.107.075138}. In the XY chain, two distinct power-law relaxation behaviors have been identified that the relaxation behavior is $\delta C_{mn}(t)\sim t^{-3/2}$ for the quench to the commensurate phase, and $\delta C_{mn}(t)\sim t^{-1/2}$ for the quench to the incommensurate phase \cite{PRB.2022.105.054301}. However, a recent article \cite{PhysRevB.2023.108.014303} find that the scaling behavior of $\delta C_{mn}(t)$ may be $\sim t^{-1}$, when the quench is from the critical point (the external field $h_{c}=1$) of the Ising transition.

It is well established that additional interactions can lead to different ground state configurations in the XY chain, which in turn have important implications for various properties. One example is the Dzyaloshinskii-Moriya (DM) interaction, an antisymmetric spin-exchange interaction that plays a crucial role in inducing antiferromagnetic \cite{JPhysChemSolid.1958.4.241, PhysRev.1960.120.91}. The DM interaction induces the emergence of a gapless phase in the XY chain \cite{PhysRevB.2008.78.214414, PhysRevA.2011.83.052112, cpb.2013.22.090313}. In this gapless phase, the ground state of the system corresponds to the configuration where all the states with $\varepsilon_{k}<0$ are filled and $\varepsilon_{k}>0$ are empty. The gapless phase has significant implications for various properties of the quantum system, such as quantum phase transitions \cite{cpb.2013.22.090313, PRE2020.102.032127, EPJB.2020.93.80, JPCM2021.33.295401}, nonequilibrium thermodynamics \cite{PRE2018.98.022107}, dynamical quantum phase transitions \cite{JPCM2018.30.42LT01, CPB2022.31.060505}, quantum speed limit \cite{PRA2023.107.042427}, and others \cite{PRB2022.105.L060401}. Therefore, it is highly intriguing to study the impact of the gapless phase on the dynamical relaxation behavior.

In this paper, we study the dynamical relaxation behavior of $C_{mn}(t)$ in the extended XY model with the gapless phase, where the gapless phase is induced by the additional interaction: the DM interaction and the XZY-YZX type of three-site interaction. For a quench from the gapped phase, we find that the dynamical relaxation behavior is not affected by the additional interaction. This is due to that in both cases, the excitation spectrum satisfies $\varepsilon_{k}+\varepsilon_{-k}=2\omega_{k}$, where $\omega_{k}$ is exactly the excitation spectrum of the XY chain without the additional interaction. However, for the quench from the gapless phase, we find that the scaling behavior of $\delta C_{mn}(t)$ is changed to $\sim t^{-1}$ for the quench to the commensurate phase, and the decay of $\delta C_{mn}(t)$ follows $\sim t^{-1}$ or $\sim t^{-1/2}$ for the quench to the incommensurate phase depending on the parameters of pre-quench Hamiltonian. This change in the scaling behavior can be attributed to the broken inverse symmetry of the excitation spectrum $(\varepsilon_{k}\neq\varepsilon_{-k})$ induced by the additional interaction. Consequently, the ground state of the pre-quench Hamiltonian in the gapless phase contains the single-occupied quasiparticle states, which do not contribute to $\delta C_{mn}(t)$.

The paper is organized as follows. In Sec.~\uppercase\expandafter {\romannumeral2}, we introduce the general expression of the XY chain with gapless phase, and give the formula of $C_{mn}(t)$ for various initial ground state. In Secs.~\uppercase\expandafter {\romannumeral3} and \uppercase\expandafter {\romannumeral4}, we consider the dynamical relaxation behaviors in the XY chain with the DM interaction and the XZY-YZX type of three-site interactions, for which the inverse symmetry of the excitation spectrum is broken. All possible quench protocols are considered. In Sec.~\uppercase\expandafter {\romannumeral5}, we discuss the results in the quench from the XX line of the XY model, for which the excitation spectrum satisfies the inverse symmetry with respect to $k=0$. In Sec.~\uppercase\expandafter {\romannumeral6}, we summary our results and conclude comments for the dynamical relaxation behavior in the XY chain with gapless phase.

\section{Models}

The Hamiltonian for the extended XY chain can be expressed by
\begin{equation}\label{extend.XY.Hamiltonian}
  \begin{split}
    H & = H_{XY} + H_{ex} \\
      & = -\frac{1}{2}\sum_{n=1}^{N}(\frac{1+\gamma}{2}\sigma_{n}^{x}\sigma_{n+1}^{x}+\frac{1-\gamma}{2}\sigma_{n}^{y}\sigma_{n+1}^{y}+h\sigma_{n}^{z}) \\
      & \quad\quad + H_{ex},
  \end{split}
\end{equation}
where $\sigma_{n}^{x,y,z}$ are the Pauli operators defined on the lattice site $n$, $\gamma$ represents the anisotropic parameter, $h$ denotes the external magnetic field. $H_{ex}$ denotes the additional interaction inducing the gapless phase, given by
\begin{equation}\label{additional.term}
  \begin{split}
    H_{ex} & = -\frac{1}{2}\sum_{n=1}^{N}[D(\sigma_{n}^{x}\sigma_{n+1}^{y} - \sigma_{n}^{y}\sigma_{n+1}^{x}) \\
      & \quad\quad +F(\sigma_{n-1}^{x}\sigma_{n}^{z}\sigma_{n+1}^{y} - \sigma_{n-1}^{y}\sigma_{n}^{z}\sigma_{n+1}^{x})],
  \end{split}
\end{equation}
where $D$ and $F$ denote the strength of nearest-neighbor and next-nearest-neighbor off-diagonal exchange interaction. When $F=0$, $H_{ex}$ reduces to the DM interaction, which describes an antisymmetric interaction \cite{JPhysChemSolid.1958.4.241, PhysRev.1960.120.91}. On the other hand, when $D=0$, $H_{ex}$ describes the next-nearest-neighbor hopping through the XZY-YZX type of three-spin interaction, which introduces gapless phases in the anisotropic XY chain \cite{JSM2012.P01003, PcB2015.463.1, PRB2016.93.214417}.

In this paper, we impose the periodic boundary conditions with $\sigma_{N+1}=\sigma_{1}$. By implementing the Jordan-Wigner transformation, the Hamiltonian (\ref{extend.XY.Hamiltonian}) can be written as a quadratic form of the spinless fermion model \cite{Suzuki2013}:
\begin{equation}
  H = \sum_{mn} c_{m}^{\dag}A_{mn}c_{n} + \frac{1}{2}\sum_{mn}(c_{m}^{\dag}B_{mn}c_{n}^{\dag}+h.c.),
\end{equation}
where $c_{n}$ and $c_{n}^{\dag}$ are fermion annihilation and creation operators, respectively. By applying the Fourier transformation, the Hamiltonian is written in momentum space as
\begin{equation}
  H = \sum_{k>0}\Psi_{k}^{\dag}\mathbb{H}_{k}\Psi_{k},
\end{equation}
where $\Psi_{k}=(c_{k}, c_{-k}^{\dag})^{T}$ are Nambu spinors, and $\mathbb{H}_{k}$ are associated Bloch Hamiltonian. The Hamiltonian can be further expressed as the diagonal form
\begin{equation}\label{diagonal.Hamil}
    H = \sum_{k>0}H_{k} = \sum_{k>0} [\varepsilon_{k}(\eta_{k}^{\dag}\eta_{k}-\frac{1}{2}) + \varepsilon_{-k}(\eta_{-k}^{\dag}\eta_{-k}-\frac{1}{2})]
\end{equation}
after using the Bogoliubov transformation $\eta_{k} = \cos{\theta_{k}}c_{k}+i\sin{\theta_{k}}c_{-k}^{\dag}$. Here, $\theta_{k}$ is the Bogoliubov angle. We consider all possible ground state configuration, in which the ground state is related to the quasiparticle excitation spectrum $\varepsilon_{k}$, that is \cite{PRE2015.91.032137, CPB2022.31.060505}
\begin{equation}
  \begin{split}
    |G\rangle & = \bigotimes_{k>0}|G\rangle_{k}, \\
      & |G\rangle_{k} = \left\{
      \begin{array}{lr}
        |0_{k}0_{-k}\rangle, & \varepsilon_{k},\varepsilon_{-k}>0, \\
        |0_{k}1_{-k}\rangle, & \varepsilon_{k}>0,\varepsilon_{-k}\leq0, \\
        |1_{k}0_{-k}\rangle, & \varepsilon_{k}\leq0,\varepsilon_{-k}>0, \\
        |1_{k}1_{-k}\rangle, & \varepsilon_{k},\varepsilon_{-k}\leq0.
      \end{array}
    \right.
  \end{split}
\end{equation}

In a quench protocol, the initial state of the system is prepared in the ground state of $H(h_{0},\gamma_{0})$, i.e. $|\psi_{0}\rangle = |G\rangle$. At $t>0$, the Hamiltonian parameters are suddenly changed to $(h_{1},\gamma_{1})$, and the system is driven by the time-evolution operator $U(t)=e^{-i\tilde{H}t}=e^{-iH(h_{1},\gamma_{1})t}$. The time-evolved state at the arbitrary time is then given by
\begin{equation}\label{time-evolved.state}
    |\psi(t)\rangle = e^{-i\tilde{H}t}|\psi_{0}\rangle = \bigotimes_{k>0} e^{-i\tilde{H}_{k}t}|G\rangle_{k},
\end{equation}
where $|G\rangle_{k}$ is not the eigenstate of the post-quench Hamiltonian $\tilde{H}$. Considering the quasiparticle operators between the pre- and post-quench Hamiltonian are related by the Bogoliubov transformation $\eta_{k} = \cos{\alpha_{k}}\tilde{\eta}_{k} - i\sin{\alpha_{k}}\tilde{\eta}_{-k}^{\dag}$ with $\alpha_{k} = \theta_{k}-\tilde{\theta}_{k}$, we obtain the eigenstates of the pre-quench Hamiltonian $H_{k}$ as a superposition of eigenstates of $\tilde{H}_{k}$ by
\begin{widetext}
\begin{equation}\label{eigenstate.relation}
    \left\{
      \begin{array}{l}
        |0_{k}0_{-k}\rangle = \cos{\alpha_{k}}|\tilde{0}_{k}\tilde{0}_{-k}\rangle - i\sin{\alpha_{k}}|\tilde{1}_{k}\tilde{1}_{-k}\rangle, \\
        |0_{k}1_{-k}\rangle = |\tilde{0}_{k}\tilde{1}_{-k}\rangle, \\
        |1_{k}0_{-k}\rangle = |\tilde{1}_{k}\tilde{0}_{-k}\rangle, \\
        |1_{k}1_{-k}\rangle = -i\sin{\alpha_{k}}|\tilde{0}_{k}\tilde{0}_{-k}\rangle + \cos{\alpha_{k}}|\tilde{1}_{k}\tilde{1}_{-k}\rangle.
      \end{array}
    \right.
\end{equation}
Then the time-evolved state is given by
\begin{equation}\label{time-evolved.state.k}
    |\psi_{k}(t)\rangle = e^{-i\tilde{H}_{k}t}|G\rangle_{k} = \left\{
      \begin{array}{l}
        \cos{\alpha_{k}}e^{i(\tilde{\varepsilon}_{k}+\tilde{\varepsilon}_{-k})t/2}|\tilde{0}_{k}\tilde{0}_{-k}\rangle-i\sin{\alpha_{k}}e^{-i(\tilde{\varepsilon}_{k}+\tilde{\varepsilon}_{-k})t/2}|\tilde{1}_{k}\tilde{1}_{-k}\rangle, \quad \varepsilon_{k},\varepsilon_{-k}>0, \\
        e^{i(\tilde{\varepsilon}_{k}-\tilde{\varepsilon}_{-k})t/2}|\tilde{0}_{k}\tilde{1}_{-k}\rangle, \quad \varepsilon_{k}>0,\varepsilon_{-k}\leq0, \\
        e^{i(-\tilde{\varepsilon}_{k}+\tilde{\varepsilon}_{-k})t/2}|\tilde{1}_{k}\tilde{0}_{-k}\rangle, \quad \varepsilon_{k}\leq0,\varepsilon_{-k}>0, \\
        -i\sin{\alpha_{k}}e^{i(\tilde{\varepsilon}_{k}+\tilde{\varepsilon}_{-k})t/2}|\tilde{0}_{k}\tilde{0}_{-k}\rangle + \cos{\alpha_{k}}e^{-i(\tilde{\varepsilon}_{k}+\tilde{\varepsilon}_{-k})t/2}|\tilde{1}_{k}\tilde{1}_{-k}\rangle, \quad \varepsilon_{k},\varepsilon_{-k}\leq0.
      \end{array}
    \right.
\end{equation}
\end{widetext}

To observe the dynamical relaxation behavior following the quench, we investigate the fermionic two-point correlation functions $C_{mn}(t) = \langle\psi(t)|c_{m}^{\dag}c_{n}|\psi(t)\rangle$ following the Refs.~\onlinecite{PRB.2022.105.054301, PhysRevB.2023.108.014303}. By considering the various configurations of the ground states, we obtain the difference $\delta C_{mn}(t)$ of the two-point correlation function from its steady-state values for a long time by
\begin{equation}
      \delta C_{mn}(t) = C_{mn}(t)-C_{mn}(\infty)= \int_{0}^{\pi}\frac{dk}{2\pi}\delta C_{mn}^{k}(t),
\end{equation}
where every component $\delta C_{mn}^{k}(t)$ is dependent on the initial states, i.e.
\begin{widetext}
\begin{equation}\label{eq:delta.Cmn(t)}
  \delta C_{mn}^{k}(t) = \left\{
    \begin{array}{cr}
      \sin{2\tilde{\theta}_{k}}\sin{2\alpha_{k}}\cos{[(\tilde{\varepsilon}_{k}+\tilde{\varepsilon}_{-k})t]}\cos{[k(n-m)]}, & \varepsilon_{k},\varepsilon_{-k}>0, \\
      0, & \varepsilon_{k}>0, \varepsilon_{k}<0, \\
      0, & \varepsilon_{k}<0, \varepsilon_{k}>0, \\
      \sin{2\tilde{\theta}_{k}}\sin{2\alpha_{k}}\cos{[(\tilde{\varepsilon}_{k}+\tilde{\varepsilon}_{-k})t]}\cos{[k(n-m)]}, & \varepsilon_{k},\varepsilon_{-k}<0.
    \end{array}
  \right.
\end{equation}
\end{widetext}
Eq.~(\ref{eq:delta.Cmn(t)}) indicates that the single-occupied quasiparticle initial states $|1_{k},0_{-k}\rangle$ and $|0_{k},1_{-k}\rangle$ do not contribute to $\delta C_{mn}(t)$.

\section{Extended XY chain with DM interaction}
Now we consider the extended XY chain with the DM interaction, in which the Hamiltonian is given by (\ref{extend.XY.Hamiltonian}) with $F=0$. By using the Jordan-Wigner and Bogoliubov transformations, the system can be expressed as the diagonal form (\ref{diagonal.Hamil}) with the quasiparticle excitation spectrum
\begin{equation}\label{eq:energy.spectrum.dm}
    \varepsilon_{k} = -2D\sin{k}+\omega_{k},
\end{equation}
where $\omega_{k} = \sqrt{(h+\cos{k})^{2}+\gamma^{2}\sin^{2}{k}}$. The Bogoliubov angles satisfy
\begin{eqnarray}
    u_{k} &=& \cos{\theta_{k}} = \frac{h+\cos{k}-\omega_{k}}{\sqrt{2[\omega_{k}^{2} - (h+\cos{k})\omega_{k}]}}, \\
    v_{k} &=& \sin{\theta_{k}} = \frac{\gamma\sin{k}}{\sqrt{2[\omega_{k}^{2} - (h+\cos{k})\omega_{k}]}}.
\end{eqnarray}
It should be noticed that the Bogoliubov angles are independent on the strength of DM interaction.

\begin{figure}[t]
    \centering
    \includegraphics[width=\linewidth]{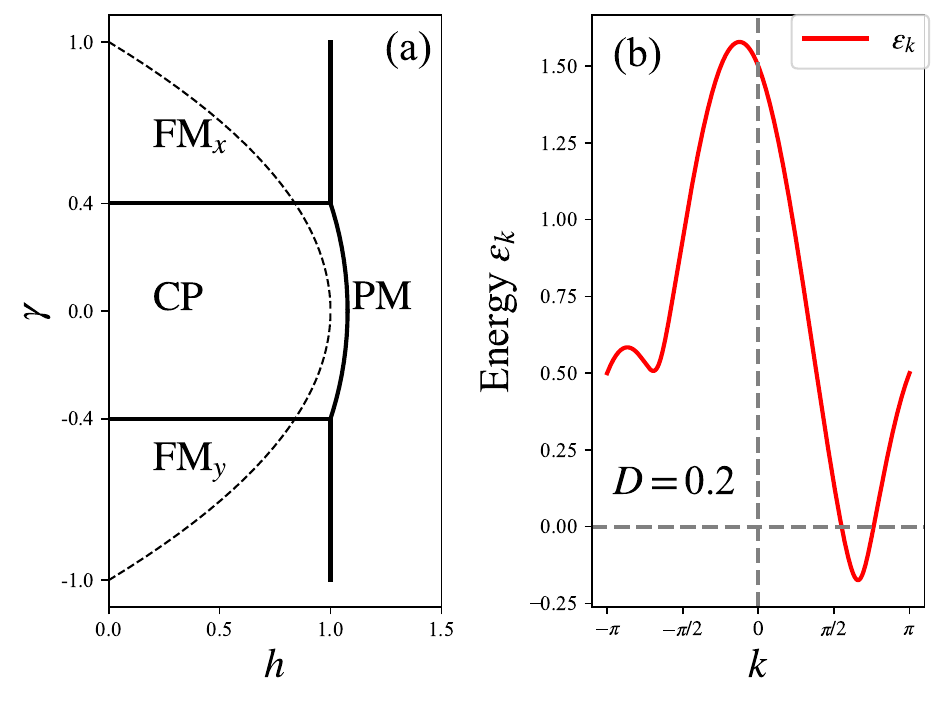}
    \caption{(a) The phase diagram of the extended XY chain with the DM interaction for $D=0.2$. The solid line between the CP and PM phases corresponds to $h=\sqrt{4D^{2}-\gamma^2+1}$. The dashed line denotes the critical lines between the commensurate and incommensurate phases, corresponding to $h = 1-\gamma^{2}$, known as disorder line. (b) The energy spectra for $(h=0.5,\gamma=0.2)$ in the CP phase. The energy spectra do not satisfy the inverse symmetry, i.e. $\varepsilon_{k}\neq\varepsilon_{-k}$.}
    \label{fig:phase-dm}
\end{figure}

Fig.~\ref{fig:phase-dm}~(a) displays the phase diagram of the extended XY chain with DM interaction for $D=0.2$. The phase diagram consists of four parts: the ferromagnetic phase along $x-$direction (FM$_{x}$), the paramagnetic phase (PM), the ferromagnetic phase along $y-$direction (FM$_{y}$), and the chiral gapless phase (CP). The dashed line denotes the critical lines between the commensurate and incommensurate phases, corresponding to $h = 1-\gamma^{2}$, known as disorder line (DL). The FM$_{x}$, FM$_{y}$ and PM phases are the gapped phases, in which the ground state is
\begin{equation}
  |G\rangle = \bigotimes_{k\in(0,\pi]}|0_{k}0_{-k}\rangle.
\end{equation}
The CP phase is the gapless phase, in which the ground state is
\begin{equation}
  |G\rangle = \bigotimes_{k_{1}>0}|0_{k_{1}}0_{-k_{1}}\rangle \bigotimes_{k_{2}>0}|1_{k_{2}}0_{-k_{2}}\rangle
\end{equation}
with $\varepsilon_{k_{1}}>0$ and $\varepsilon_{k_{2}}<0$.

From Eq.~(\ref{eq:delta.Cmn(t)}), the $\delta C_{mn}(t)$ in the XY chain with DM interaction is given by
\begin{equation}\label{eq: delta_C_dm}
  \delta C_{mn}(t) = \left\{
                        \begin{array}{ll}
                          \int_{k\in(0,\pi]}\frac{dk}{2\pi}\delta C_{mn}^{k}(t), & \text{from gapped phase}, \\
                          \int_{k\in\{k_{1}\}}\frac{dk}{2\pi}\delta C_{mn}^{k}(t), & \text{from gapless phase},
                        \end{array}
                     \right.
\end{equation}
with
\begin{equation}
  \delta C_{mn}^{k}(t) = \sin{2\tilde{\theta}_{k}}\sin{2\alpha_{k}}\cos{(2\tilde{\omega}_{k}t)}\cos{[k(m-n)]}.
\end{equation}
It should be noticed that here we have $\tilde{\varepsilon}_{k}+\tilde{\varepsilon}_{-k}=2\tilde{\omega}_{k}$, and $\tilde{\omega}_{k}$ is exactly the excitation spectrum of the XY chain without the DM interaction.

\subsection{Quench from gapped phases}

\begin{figure}
    \centering
    \includegraphics[width=\linewidth]{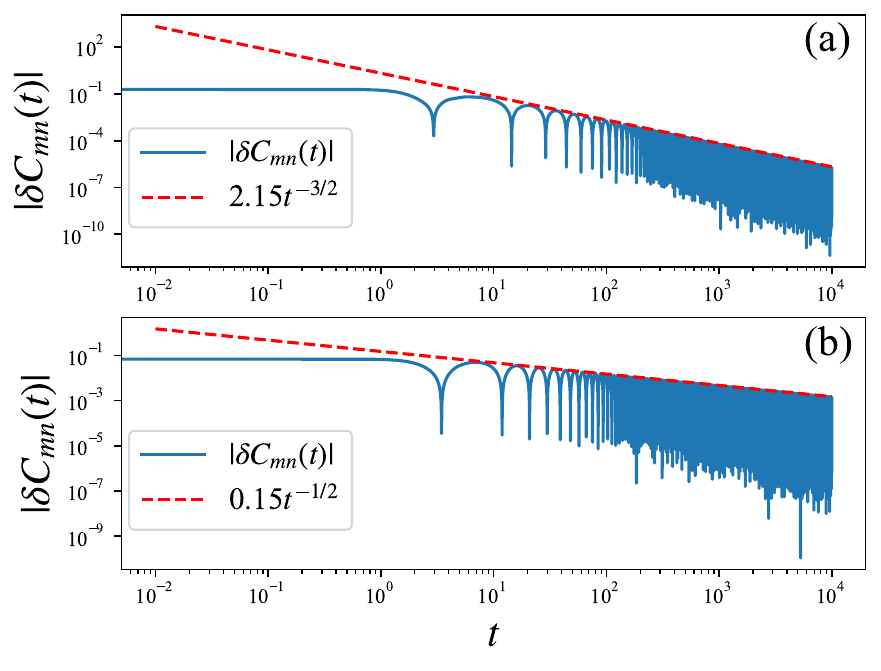}
    \caption{(a) $|\delta C_{mn}(t)|$ as a function of $t$ for a quench from the PM phase to the commensurate phase, which is from $h_{0}=100$ to $h_{1}=0.9$ with fixed $\gamma_{0}=\gamma_{1}=0.5$.  (b) $|\delta C_{mn}(t)|$ for a quench from the PM phase to the incommensurate phase, which is from $h_{0}=100$ to $h_{1}=0.5$ with fixed $\gamma_{0}=\gamma_{1}=0.2$.}
    \label{fig:delta_C_gapped}
\end{figure}

Firstly, we consider the quench protocols from the gapped phase. In Fig.~\ref{fig:delta_C_gapped}, we display the $\delta C_{mn}(t)$ as a function of $t$ for the quench from the PM phase to both the commensurate and incommensurate phases, where $|n-m|=1$. It is evident that for the quench from the gapped phase to the commensurate phase, $\delta C_{mn}(t)$ exhibits a scaling behavior of $\sim t^{-3/2}$, while for the quench to the incommensurate phase, the scaling behavior is given by $\delta C_{mn}(t)\sim t^{-1/2}$.

The relaxation behavior of $\delta C_{mn}(t)$ can be explained by the method of stationary point approximation. For the quench from the gapped phase, we have
\begin{equation}\label{eq:complex_delta_C}
    \delta C_{mn}(t) = \mathrm{Re}[I(t)],
\end{equation}
where Euler's formula is used to obtain $(|n-m|=1)$
\begin{equation}\label{eq:saddle-form}
    \begin{split}
        I(t) & = \frac{1}{2\pi}\int_{0}^{\pi} dk  \sin{2\tilde{\theta}_{k}}\sin{2\alpha_{k}}e^{2i\tilde{\omega}_{k}t}\cos{k} \\
             & = \frac{1}{2\pi}\int_{0}^{\pi} dk f(k)e^{ig(k)t}
    \end{split}
\end{equation}
with $f(k) = \sin{2\tilde{\theta}_{k}}\sin{2\alpha_{k}}$ and $g(k) = 2\tilde{\omega}_{k}$. The exponential term in $I(t)$ oscillates rapidly, so that $I(t)$ is determined by the integrals around the stationary points $k_{0}$, which satisfy $g'(k_{0})=0$. It also should be noticed that in the XY chain with DM interaction, the stationary points is given by $\frac{\partial\tilde{\omega}_{k}}{\partial k}=0$, which is independent of the DM interaction.

Specifically, for the quench to the commensurate phase, there are two stationary points $k=0, \pi$. The contributions for the integrals around stationary points $k=0, \pi$ to $I(t)$ both have the approximate behavior of $\sim t^{-3/2}$ [see Appendix~A]. Consequently, the relaxation behavior of $\delta C_{mn}(t)$ follows the scaling behavior $t^{-3/2}$ for the long time for the quench from the gapped phase to the commensurate phase. However, for the quench to the incommensurate phase, there is an extra stationary point $k_{m}$, corresponding to the minimum value of $\tilde{\omega}_{k}$, besides two stationary points $k=0, \pi$. The integral around $k_{m}$ contributes a slower scaling decay $\sim t^{-1/2}$, than that of $k=0, \pi$. Therefore, the relaxation behavior of $\delta C_{mn}(t)$ dominates for the scaling behavior $\sim t^{-1/2}$, when quenching from the gapped phase to the incommensurate phase.

The dynamical relaxation behavior in the quench protocols from the gapped phase is only determined by whether the post-quench Hamiltonian is in the commensurate or incommensurate phase. This is similar to the behavior observed in the XY chain, which suggests that the DM interaction does not affect the relaxation behavior of $\delta C_{mn}(t)$. The reason can be explained by that the excitation spectrum of the XY chain with DM interaction satisfies $\varepsilon_{k}+\varepsilon_{-k}=2\omega_{k}$, where $\omega_{k}$ is exactly the excitation spectrum of the XY chain.

\subsection{Quench from gapless phase}

\begin{figure}[t]
    \centering
    \includegraphics[width=\linewidth]{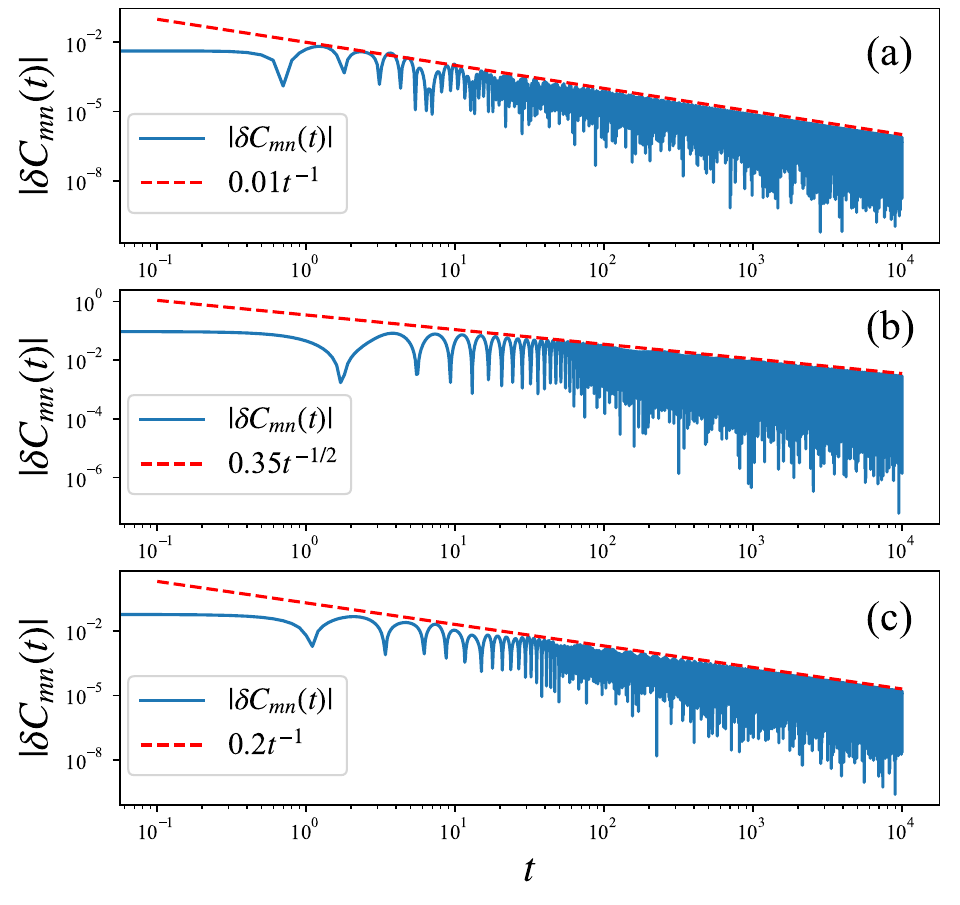}
    \caption{(a) $|\delta C_{mn}(t)|$ as a function of $t$ for a quench from the CP phase to the commensurate phase, from $h_{0}=0.2$ to $h_{1}=2.0$ with fixed $\gamma_{0}=\gamma_{1}=0.2$.  $|\delta C_{mn}(t)|$ for a quench from the CP phase to the incommensurate part of FM$_{x}$ phase, (b) from $(h_{0}=0.2,\gamma_{0}=0.1)$ to $(h_{1}=0.5,\gamma_{1}=0.5)$, and (c) from $(h_{0}=0.2,\gamma_{0}=0.1)$ to $(h_{1}=0.2,\gamma_{1}=0.5)$.}
    \label{fig:delta_C_gapless}
\end{figure}

Now, we consider the quench protocols from the gapless phase. In Fig.~\ref{fig:delta_C_gapless}, we display the $\delta C_{mn}(t)$ as a function of $t$ for the quench from the gapless chiral phase to both the commensurate PM phase and incommensurate part of the FM$_x$ phases. It can be observed that for the quench from the gapless phase to the commensurate PM phase, the scaling behavior is given by $\delta C_{mn}(t)\sim t^{-1}$, and for the quench to the incommensurate phase, the scaling behavior is $\delta C_{mn}(t)\sim t^{-1/2}$ or $\sim t^{-1}$.

\begin{figure}[t]
    \centering
    \includegraphics[width=\linewidth]{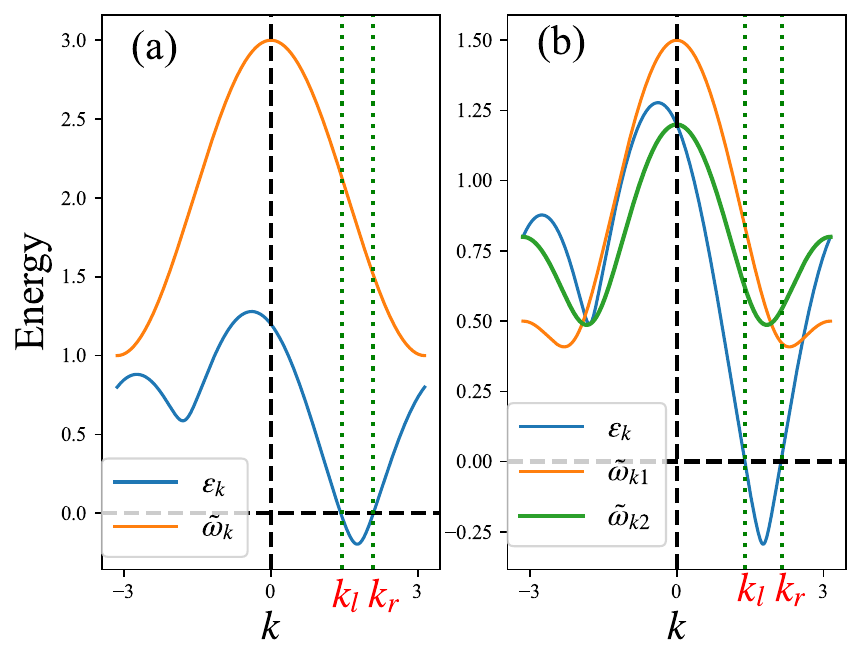}
    \caption{(a) Energy spectra $\varepsilon_{k}$ and $\tilde{\omega}_{k}$ for the pre-quench Hamiltonian parameter $(h_{0}=0.2, \gamma_{0}=0.2)$ and the post-quench Hamiltonian parameter $(h_{1}=2.0, \gamma_{1}=0.2)$. (b) Energy spectra $\varepsilon_{k}$, $\tilde{\omega}_{k1}$ and $\tilde{\omega}_{k2}$ for the pre-quench Hamiltonian parameter $(h_{0}=0.5, \gamma_{0}=0.1)$, the post-quench Hamiltonian parameters $(h_{1}=0.5, \gamma_{1}=0.5)$, and $(h_{1}=0.2, \gamma_{1}=0.5)$. The interval $[k_{l}, k_{r}]$ does not contain the minimum value of $\tilde{\omega}_{k1}$, but contain the minimum value of $\tilde{\omega}_{k2}$. }
    \label{fig:energy.quench.dm}
\end{figure}

To explain the relaxation behavior in the quench protocol from the gapless phase, we can express the function $I(t)$ by
\begin{equation}
    I(t) = \frac{1}{2\pi}(\int_{0}^{k_{l}} + \int_{k_{r}}^{\pi})dkf(k)e^{ig(k)t},
\end{equation}
where $k_{l}, k_{r}$ are two boundary points, and $\varepsilon_{k}<0$ for $k_{l}<k<k_{r}$ [see Fig.~\ref{fig:energy.quench.dm}]. In this case, the asymptotic behavior of $I(t)$ is determined by the competition between the integrals around stationary points and the boundary points.

Specifically, for the quench from the gapless phase to the commensurate phase, there are two stationary points $k=0, \pi$ and two boundary points $k_{l}, k_{r}$ [see Fig.~\ref{fig:energy.quench.dm}~(a)]. It is already known that the stationary points $k=0, \pi$ contribute the scaling decay $\sim t^{-3/2}$. While for the boundary points, according to the generalized Riemann-Lebesgue lemma, the integral around two boundary points in the limited intervals $k\in[0,k_{l}], [k_{r},\pi]$ is given by
\begin{equation}
  \sim f(k_{l})\frac{e^{2it\tilde{\omega}_{k_{l}}}}{i\tilde{\omega}_{k_{l}}^{'}}t^{-1} + f(k_{r})\frac{e^{2it\tilde{\omega}_{k_{r}}}}{i\tilde{\omega}_{k_{r}}^{'}}t^{-1}.
\end{equation}
Therefore, the long-time scaling behavior of the integral around the boundary points is $\sim t^{-1}$. As a result, for the quench from the gapless phase to the commensurate phase, the relaxation behavior of $\delta C_{mn}(t)$ follows the slower $\sim t^{-1}$.

While for the quench from the gapless phase to the incommensurate phase, there are two different cases. The first one is that the interval $[k_{l},k_{r}]$ does not contains the minimum points $k_{m}$ [see the orange line in Fig.~\ref{fig:energy.quench.dm}~(b)]. In this case, the asymptotic behavior of $I(t)$ is determined by competition between the integrals around three stationary points $k=0, \pi, k_{m}$, and two boundary points $k_{l}, k_{r}$. It is evident that the integral around the minimum point $k_{m}$ contributes the slowest decay $\sim t^{-1/2}$. Consequently, the relaxation behavior of $\delta C_{mn}(t)$ dominates the scaling behavior $\sim t^{-1/2}$ for a long time.

On the other hand, if the interval $[k_{l}, k_{r}]$ contains the minimum points $k_{m}$ [see the orange line in Fig.~\ref{fig:energy.quench.dm}~(b)], the stationary point $k_{m}$ will not contribute to $I(t)$ anymore. In this case, the asymptotic behavior of $I(t)$ is determined by competition between the integrals around two stationary points $k=0, \pi$, and two boundary points $k_{l}, k_{r}$. Similar to the case from the gapless phase to the commensurate phase, the relaxation behavior of $\delta C_{mn}(t)$ follows the scaling behavior $\sim t^{-1}$.

\subsection{Quench from gapless phase to the disorder line}

\begin{figure}
  \centering
  \includegraphics[width=\linewidth]{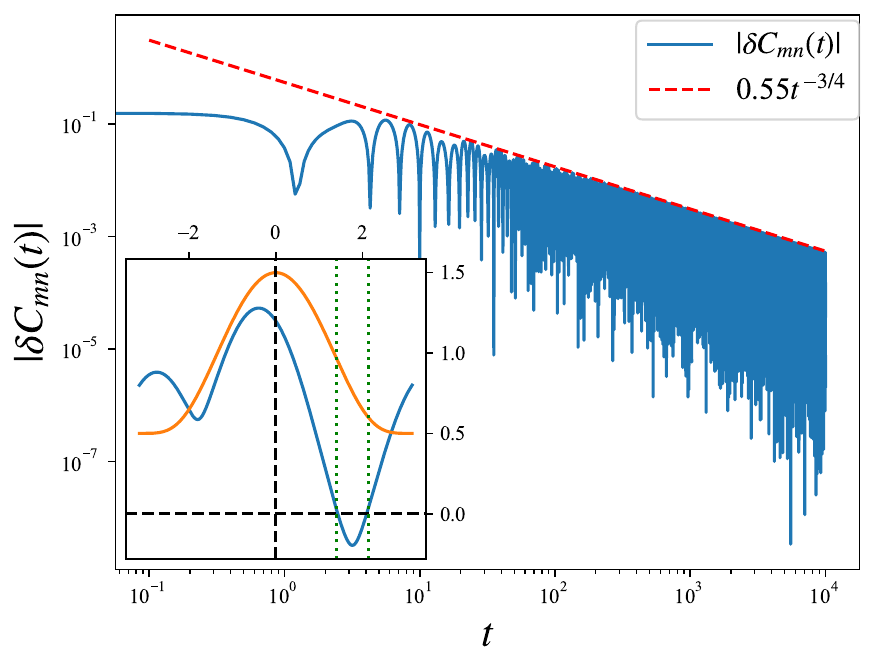}\\
  \caption{$|\delta C_{mn}(t)|$ as a function of $t$ for a quench from the gapless chiral phase to the disorder line, which is from $(h_{0}=0.2,\gamma_{0}=0.1)$ to $(h_{1}=0.5,\gamma_{1}=\frac{1}{\sqrt{2}})$. The inset graph shows energy spectra $\varepsilon_{k}$ and $\tilde{\omega}_{k}$ for the Hamiltonian parameters $(h_{0}=0.2,\gamma_{0}=0.1)$ and $(h_{1}=0.5,\gamma_{1}=\frac{1}{\sqrt{2}})$. }\label{fig:delta.C.gapless.dl}
\end{figure}

Now we consider the quench protocol from the gapless phase to the disorder line. The disorder line is the boundary between the commensurate and incommensurate phases in the XY chain. It has already been found a different relaxation behavior of $\delta C_{mn}(t)\sim t^{-3/4}$ for the quench from the gapped phase to the disorder line \cite{PRB.2022.105.054301}.  In Fig.~\ref{fig:delta.C.gapless.dl}, we display the $\delta C_{mn}(t)$ as a function of $t$ for the quench from the gapless phase to the disordered line. The relaxation behavior of $\delta C_{mn}(t)$ is observed to still follow $\sim t^{-3/4}$. To explain this, we display the excitation spectra $\varepsilon_{k}$ and $\tilde{\omega}_{k}$ in the inset graph of Fig.~\ref{fig:delta.C.gapless.dl}. The asymptotic behavior of $\delta C_{mn}(t)$ is determined by the competition between the contributions of  stationary points $k=0, \pi$ and boundary points, in which the contributions of $k=0$ and two boundary points are $\sim t^{-3/2}$ and $\sim t^{-1}$, respectively. At the stationary point $k=\pi$, we have $\frac{d\tilde{\omega}_{k}}{dk}|_{k=\pi}=\frac{d^{2}\tilde{\omega}_{k}}{dk^{2}}|_{k=\pi}=0$, corresponding to the high-order stationary point approximation. It is known that the high-order stationary point $k=\pi$ contributes the scaling decay for $\sim t^{-3/4}$, which is slower than that for $\sim t^{-3/2}$ and $\sim t^{-1}$. Consequently, the relaxation behavior of $\delta C_{mn}(t)$ follows $t^{-3/4}$ for the quench to the disorder line, regardless of whether the quench originates from the gapped or gapless phase.

\subsection{Dynamical phase diagram}

\begin{figure}
  \centering
  \includegraphics[width=0.945\linewidth]{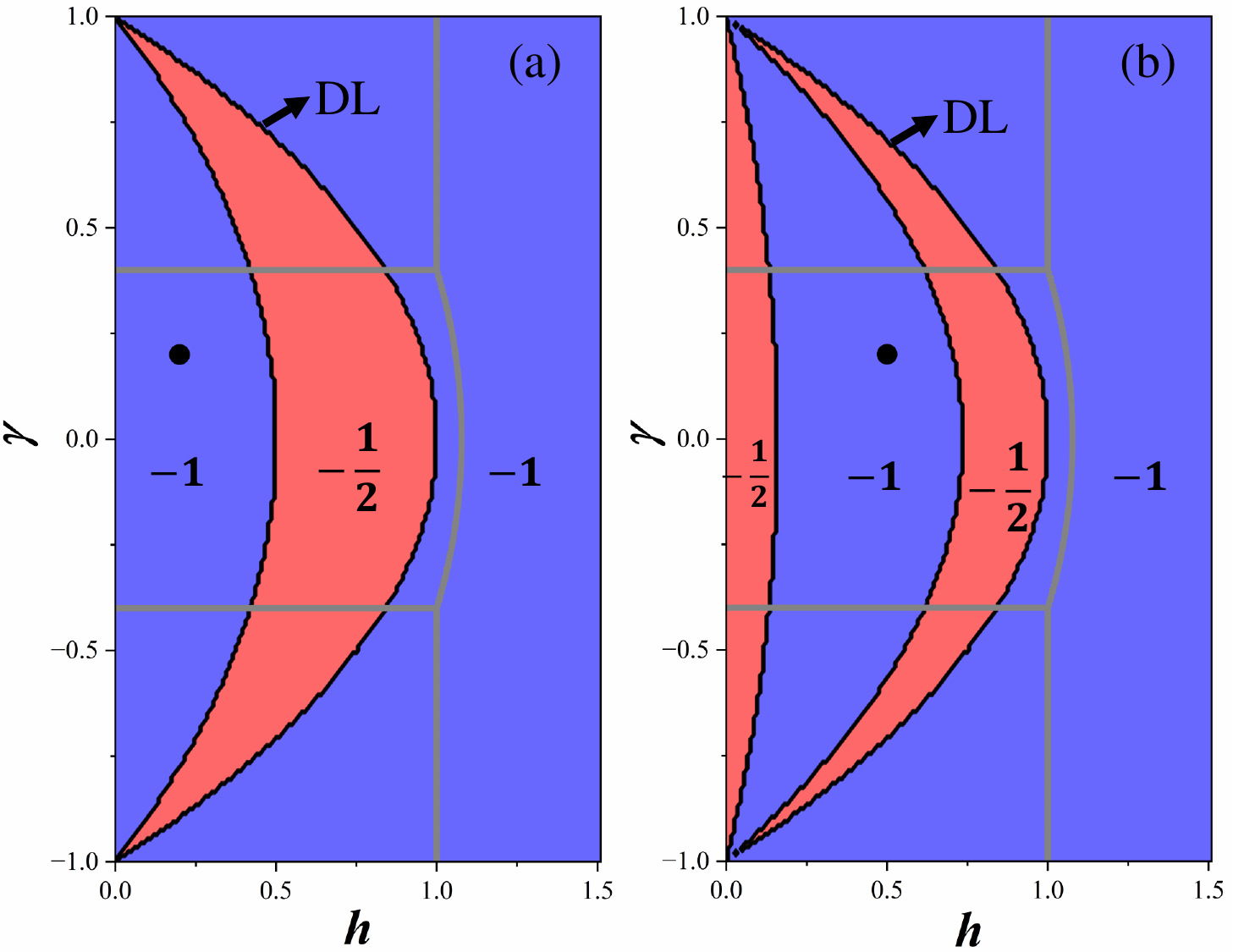}\\
  \caption{(a) The dynamical phase diagram for the quench from $(h_{0}=0.2, \gamma_{0}=0.2)$ marked by the black solid dot. (b) The dynamical phase diagram for the quench from $(h_{0}=0.5, \gamma_{0}=0.2)$. The dynamical phases are characterized by the dynamical relaxation behavior of $\delta C_{mn}(t)$, where the blue region denotes $\delta C_{mn}(t)\sim t^{-1}$, and the red region denotes $\delta C_{mn}(t)\sim t^{-1/2}$. The right boundary is exactly the disorder line. }\label{fig:dynamical.phase.diagram}
\end{figure}

In this section, we will present a schematic phase diagram that captures the different dynamical phases based on the relaxation behavior of $\delta C_{mn}(t)$. While obtaining the dynamical phase diagram for the quench protocol from the gapped phase is straightforward, as it is divided by the disorder line, we will focus on the quench protocol from the gapless phase in this discussion. The dynamical relaxation behavior of $\delta C_{mn}(t)$ is determined by the conditions of whether the post-quench Hamiltonian is in the commensurate and incommensurate phases, and whether the interval $[k_{l},k_{r}]$ contains the minimum point $k_{m}$, as discussed in previous sections. Therefore, the boundary of dynamical phases for the first condition is the disorder line, i.e. $h=1-\gamma^{2}$.

The boundary for the second condition can be obtained by
\begin{equation}\label{eq:second.boundary.dm}
  h = \frac{h_{0}\pm\sqrt{h_{0}^{2}-(1-\gamma_{0}^{2}+4D^{2})(h_{0}^{2}+\gamma_{0}^{2}-4D^{2})}}{1-\gamma_{0}^{2}+4D^{2}}(1-\gamma^{2}),
\end{equation}
where $h_{0}, \gamma_{0}$ denote the parameters of the pre-quench Hamiltonian (see Appendix~C). It is important to note that the boundary (\ref{eq:second.boundary.dm}) is dependent on the parameters of the pre-quench Hamiltonian. The coefficient in (\ref{eq:second.boundary.dm}) represents the solutions of a quadratic equation, resulting in two possible cases for the boundary for $h>0$. It is known that $1-\gamma^{2}+4D^{2}>0$ all the times, so that if $h_{0}^{2}+\gamma_{0}^{2}<4D^{2}$, we have $h_{0}<\sqrt{h_{0}^{2}-(1-\gamma_{0}^{2}+4D^{2})(h_{0}^{2}+\gamma_{0}^{2}-4D^{2})}$. In this case, there is one boundary, i.e. $h = \frac{h_{0}+\sqrt{h_{0}^{2}-(1-\gamma_{0}^{2}+4D^{2})(h_{0}^{2}+\gamma_{0}^{2}-4D^{2})}}{1-\gamma_{0}^{2}+4D^{2}}(1-\gamma^{2})$ [see Fig.~\ref{fig:dynamical.phase.diagram}~(a)]. If $h_{0}^{2}+\gamma_{0}^{2}>4D^{2}$, we have $h_{0}>\sqrt{h_{0}^{2}-(1-\gamma_{0}^{2}+4D^{2})(h_{0}^{2}+\gamma_{0}^{2}-4D^{2})}$. In this case, there two boundaries following the Eq.~(\ref{eq:second.boundary.dm}) [see Fig.~\ref{fig:dynamical.phase.diagram}~(b)].


\section{Results of the XY chain with XZY-YZX type of three-spin interaction}

\begin{figure}[t]
    \centering
    \includegraphics[width=\linewidth]{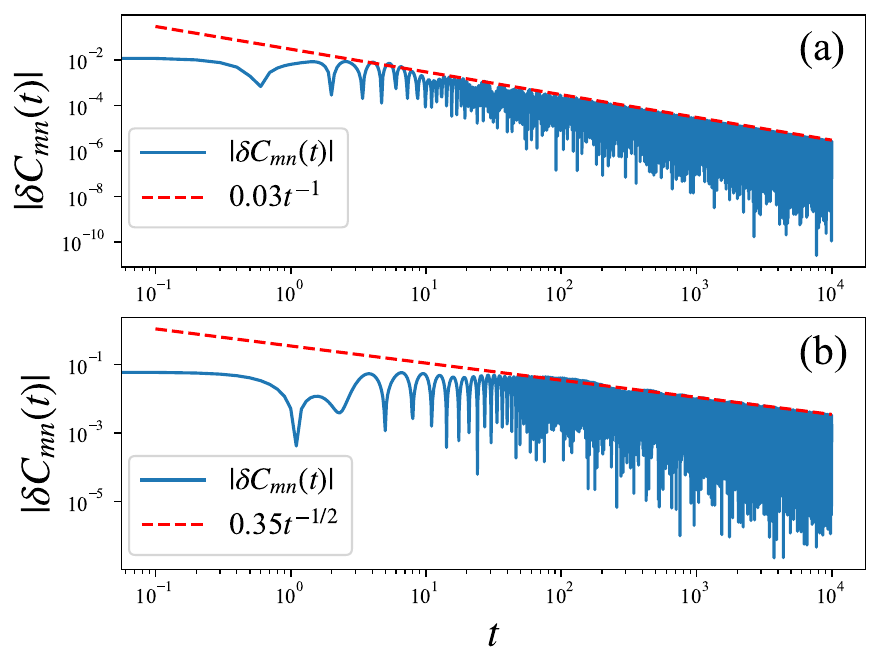}
    \caption{In the XY chain with XZY-YZX type of three-site interaction, (a) $|\delta C_{mn}(t)|$ as a function of $t$ for a quench from the gapless chiral phase to the commensurate phase, which is from $(h_{0}=0.5, \gamma_{0}=0.1)$ to $(h_{1}=2.0, \gamma_{1}=0.5)$.  (b) $|\delta C_{mn}(t)|$ for a quench from the gapless chiral phase to the incommensurate phase, which is from $(h_{0}=0.5, \gamma_{0}=0.1)$ to $(h_{1}=0.5, \gamma_{1}=0.65)$.}
    \label{fig:delta.C.XZY}
\end{figure}

\begin{figure}[t]
  \centering
  \includegraphics[width=\linewidth]{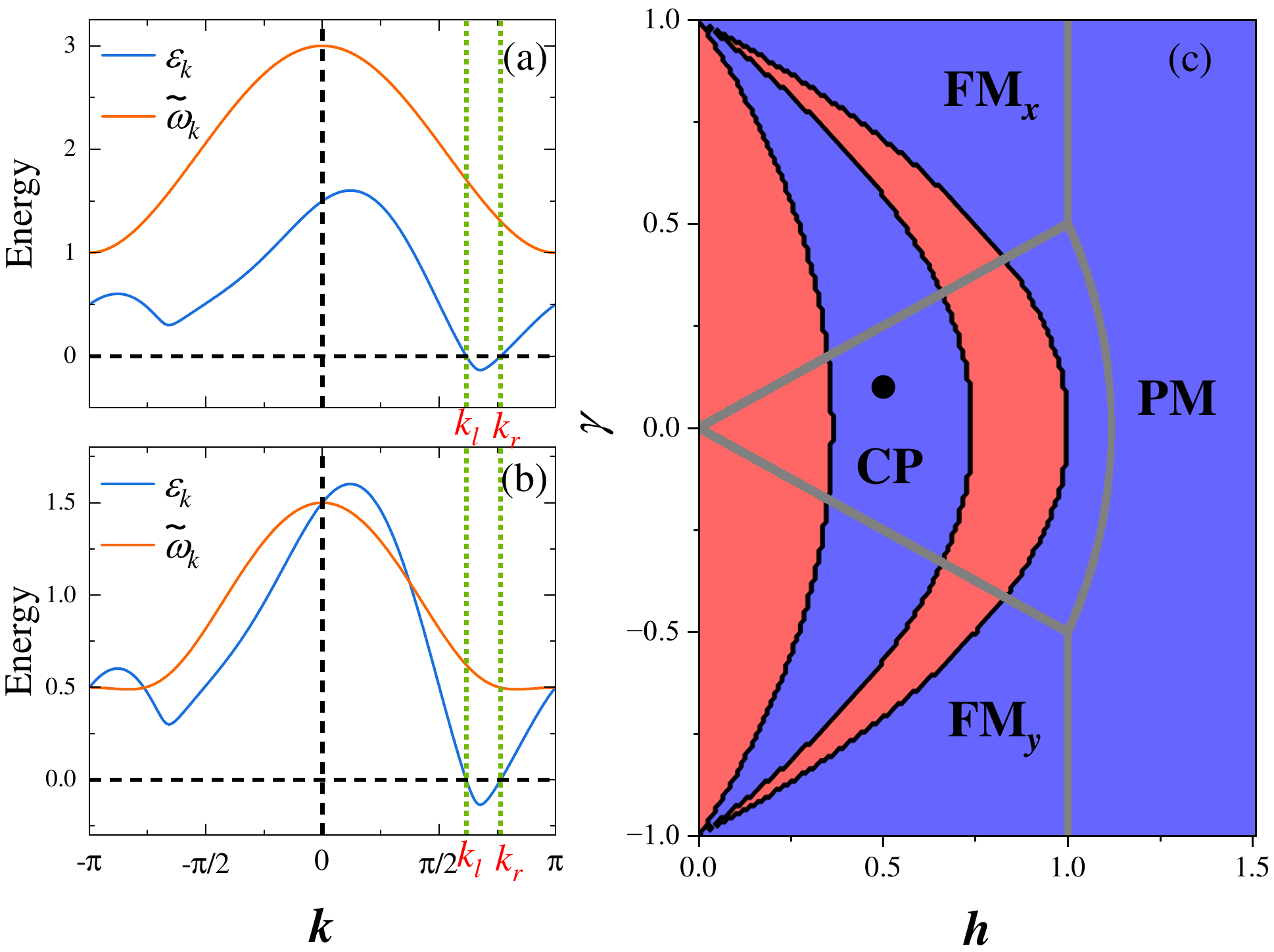}\\
  \caption{ In the XY chain with XZY-YZX type of three-site interaction, (a) energy spectra $\varepsilon_{k}$ and $\tilde{\omega}_{k}$ for the pre-quench Hamiltonian parameter $(h_{0}=0.5, \gamma_{0}=0.1)$ and the post-quench Hamiltonian parameter $(h_{1}=2.0, \gamma_{0}=0.5)$. (b) $\varepsilon_{k}$ and $\tilde{\omega}_{k}$ for the pre-quench Hamiltonian parameter $(h_{0}=0.5, \gamma_{0}=0.1)$ and the post-quench Hamiltonian parameter $(h_{1}=0.5, \gamma_{0}=0.65)$. (c) The dynamical phase diagram for the quench from $(h_{0}=0.5, \gamma_{0}=0.1)$ marked by the black solid dot. The grey solid lines are the critical lines of the quantum phase transitions. }\label{fig:dynamical.phase.diagram.XZY}
\end{figure}

Now, we consider the XY chain with XZY-YZX type of three-spin interaction, which is described by the Hamiltonian (\ref{extend.XY.Hamiltonian}) with $D=0, \beta=-1$. Similarly to the XY chain with the DM interaction, the phase diagram consists of four parts [see Fig.~\ref{fig:dynamical.phase.diagram.XZY}~(c)]: the ferromagnetic phase along $x-$direction (FM$_{x}$), the paramagnetic phase (PM), the ferromagnetic phase along $y-$direction (FM$_{y}$), and the chiral gapless phase (CP), where except the CP phase, FM$_{x}$, FM$_{y}$ and PM are the gapped phases. The quasiparticle excitation spectrum is given by \cite{JSM2012.P01003}
\begin{equation}\label{eq:energy.spectrum.XZY}
  \varepsilon_{k} = \frac{F}{2}\sin{2k} + \sqrt{(h+\cos{k})^{2} + \gamma^{2}\sin^{2}{k}},
\end{equation}
where the first term $\frac{F}{2}\sin{2k}$ breaks the inverse symmetry of the XY chain. Similar to that in the XY chain with DM interaction, the ground state in the gapped phase (FM$_x$, FM$_y$, and PM phases) is $|G\rangle=\bigotimes_{k>0}|0_{k},0_{-k}\rangle$, and in the gapless phase (CP phase) is $|G\rangle=\bigotimes_{k_{1}>0}|0_{k_{1}}0_{-k_{1}}\rangle\bigotimes_{k_{2}>0}|1_{k_{2}}0_{-k_{2}}\rangle$.

It should be noticed that similar to the case in the XY chain with DM interaction, the excitation spectrum (\ref{eq:energy.spectrum.XZY}) also satisfies $\varepsilon_{k}+\varepsilon_{-k}=2\omega_{k}$, where $\omega_{k}$ is the excitation spectrum of XY chain without the additional interaction. As a result, for the quench from the gapped phase, the XZY-YZX type of three-site interaction does not influence the relaxation behavior of $\delta C_{mn}(t)$. In the following, we show the results of quenching from the gapless CP phase.

In Fig.~\ref{fig:delta.C.XZY}, we display the $|\delta C_{mn}(t)|$ as a function of $t$ for the quench from the gapless chiral phase to both the commensurate PM phase and incommensurate part of the FM$_x$ phases, where $|n-m|=1$. It can be observed that for the quench from the gapless phase to the commensurate PM phase, the scaling behavior is given by $\delta C_{mn}(t)\sim t^{-1}$, and for the quench to the incommensurate phase, the scaling behavior is $\delta C_{mn}(t)\sim t^{-1/2}$.

Similarly to the case in the XY chain with DM interaction, the dynamical relaxation behavior can be explained by the stationary phase approximation. For the quench from the gapless chiral phase to the commensurate phase, there are two stationary points $k=0, \pi$ and two boundary points $k_{l}, k_{r}$ [see Fig.~\ref{fig:dynamical.phase.diagram.XZY}~(a)]. As mentioned before, the integrals around the stationary points at the boundary or center of Brillouin zone provide approximate behavior $\sim t^{-3/2}$, and the boundary points provide $\sim t^{-1}$. Therefore, the power-law behavior of $\delta C_{mn}(t)$ is $\sim t^{-1}$ for the quench from the chiral phase to the commensurate phase. However, for the quench from the gapless chiral phase, the asymptotic behavior depends on whether the interval $(k_{l},k_{r})$ covers the minimum value of $\tilde{\omega}_{k}$. As seen in Fig.~\ref{fig:dynamical.phase.diagram.XZY}~(b), the minimum value of $\tilde{\omega}_{k}$ is not covered in the interval $(k_{l},k_{r})$. The integral of $\delta C_{mn}(t)$ is thus contributed by three stationary points $k = 0, \pi, k_{m} (\tilde{\omega}_{k_m}=\min{\tilde{\omega}_{k}})$ and two boundary points $k_{l}, k_{r}$, where the integral around $k_{m}$ provides the slowest asymptotic decay $\sim t^{-1/2}$. Therefore, the power-law behavior of $\delta C_{mn}(t)$ is $\sim t^{-1/2}$, which agrees with the numerical results in Fig.~\ref{fig:delta.C.XZY}~(b). If the interval $(k_{l}, k_{r})$ covers the minimum value $\min{\tilde{\omega}_{k}}$, the power-law of $\delta C_{mn}(t)$ is $\sim t^{-1}$.

Finally, we obtain the dynamical phase for the quench from the point $(h_{0}=0.5, \gamma_{0}=0.1)$ [see Fig.~\ref{fig:dynamical.phase.diagram.XZY}~(c)].

\section{Results of the quench from the XX line}

\begin{figure}
    \centering
    \includegraphics[width=\linewidth]{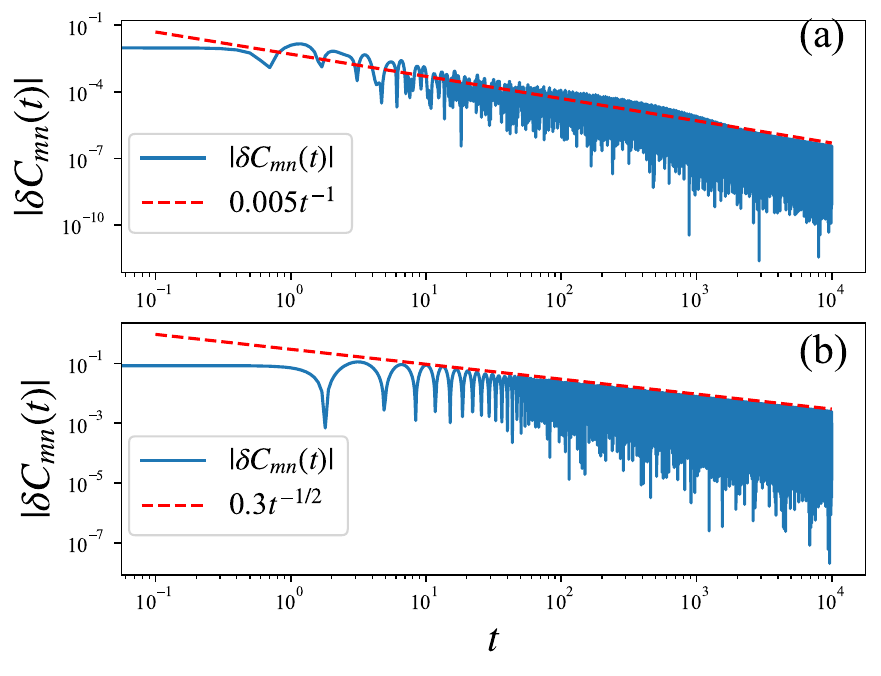}
    \caption{(a) $|\delta C_{mn}(t)|$ as a function of $t$ for a quench from the XX line to the commensurate phase, which is from $(h_{0}=0.5, \gamma_{0}=0.0001)$ to $(h_{1}=2.0, \gamma_{1}=0.5)$.  (b) $|\delta C_{mn}(t)|$ for a quench from the XX line to the incommensurate phase, which is from $(h_{0}=0.5, \gamma_{0}=0.0001)$ to $(h_{1}=0.5, \gamma_{1}=0.5)$.}
    \label{fig:delta.C.XX}
\end{figure}

In the previous sections, we discuss the dynamical relaxation behaviors in the XY chain with DM interaction ($D\neq0,F=0$) and the XZY-YZX type of three-site interaction ($D=0,F\neq0$), respectively. In both models, the energy spectra are asymmetric, so when quenching from the gapless phase, the initial state consists of the vacuum states $|0_{k}0_{-k}\rangle$ and the single-occupied state $|1_{k}0_{-k}\rangle$. Now, let us consider another special case, i.e. quench from the XX line in the XY chain ($\gamma=0,D=F=0,h\leq1$). In this case, the quasiparticle excitation spectrum $\varepsilon_{k}$ of the pre-quench Hamiltonian satisfies the inverse symmetry with respect of $k=0$, which is given by
\begin{equation}\label{energy.XX}
  \varepsilon_{k} = h + \cos{k}.
\end{equation}
To calculate the $\delta C_{mn}(t)$, we consider the XX line as the $\gamma\rightarrow0$ limit.

In Fig.~\ref{fig:delta.C.XX}, we display the $\delta C_{mn}(t)$ as a function of $t$ for the quench from the gapless chiral phase to both the commensurate PM phase and incommensurate part of the FM$_x$ phases, where $|n-m|=1$. It can be observed that for the quench from the gapless phase to the commensurate PM phase, the scaling behavior is given by $\delta C_{mn}(t)\sim t^{-1}$, and for the quench to the incommensurate phase, the scaling behavior is $\delta C_{mn}(t)\sim t^{-1/2}$.

\begin{figure}
  \centering
  \includegraphics[width=\linewidth]{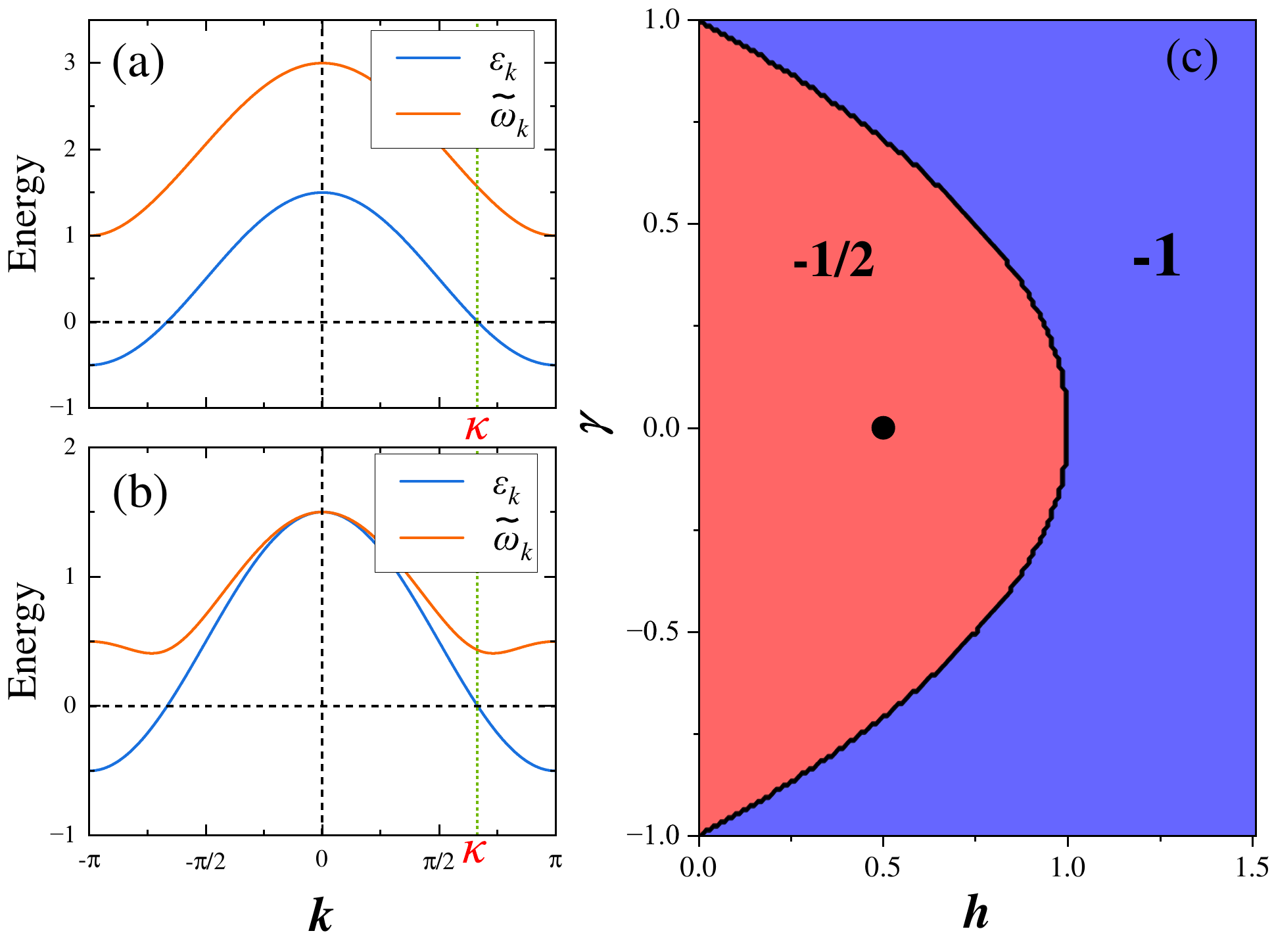}\\
  \caption{ (a) Energy spectra $\varepsilon_{k}$ and $\tilde{\omega}_{k}$ for the pre-quench Hamiltonian parameter $(h_{0}=0.5, \gamma_{0}=0.0001)$ and the post-quench Hamiltonian parameter $(h_{1}=2.0, \gamma_{0}=0.5)$. (b) Energy spectra $\varepsilon_{k}$ and $\tilde{\omega}_{k}$ for the pre-quench Hamiltonian parameter $(h_{0}=0.5, \gamma_{0}=0.0001)$ and the post-quench Hamiltonian parameter $(h_{1}=0.5, \gamma_{0}=0.5)$. (c) The dynamical phase diagram for the quench from $(h_{0}=0.5, \gamma_{0}=0.0001)$ marked by the black solid dot. }\label{fig:dynamical.phase.diagram.XX}
\end{figure}

The Eq.~\ref{energy.XX} reveals that the quasiparticle excitation spectrum of the XX case satisfies the inverse symmetry with respect to $k=0$, i.e. $\varepsilon_{k}=\varepsilon_{-k}$. The inverse symmetry guarantees the ground state of the pre-quench Hamiltonian is given by
\begin{equation}
  |G\rangle = \bigotimes_{0<k<\kappa}|0_{k}0_{-k}\rangle\bigotimes_{\kappa<k<\pi}|1_{k}1_{-k}\rangle,
\end{equation}
with $\varepsilon_{k},\varepsilon_{-k}>0$ for $k<\kappa$ and $\varepsilon_{k},\varepsilon_{-k}<0$ for $k>\kappa$ [see Fig.~\ref{fig:dynamical.phase.diagram.XX}~(a) and (b)]. The integral of $\delta C_{mn}(t)$ is thus separated as two parts, given by
\begin{equation}
  \begin{split}
    \delta C_{mn}(t) & = \frac{1}{2\pi}\int_{0}^{\kappa}dk\sin{2\tilde{\theta}_{k}}\sin{2\alpha_{k}}\cos{(2\tilde{\omega}_{k}t)}\cos{k} \\
      & -\frac{1}{2\pi}\int_{\kappa}^{\pi}dk\sin{2\tilde{\theta}_{k}}\sin{2\alpha_{k}}\cos{(2\tilde{\omega}_{k}t)}\cos{k}
  \end{split}
\end{equation}
Therefore, the relaxation behavior of $\delta C_{mn}(t)$ is determined by the integrals around the boundary point $\kappa$ and the stationary points $k=0,\pi$ and $k=k_{m}$, where $k_{m}$ corresponds to the minimum value of $\tilde{\omega}_{k}$ for the system in the incommensurate phase. It also should be noticed that the case of critical quench in Ref.~\onlinecite{PhysRevB.2023.108.014303} can be treated as an exceptional case of our theory, in which the zero excitation spectrum $\varepsilon_{k}=0$ is located at the center or boundary of the Brillouin zone, i.e. $\kappa=0,\pi$.

Specifically, for the quench from the XX line to the commensurate phase, the $\delta C_{mn}(t)$ is $\sim at^{-3/2}+bt^{-1}$, which dominates for the scaling behavior of $t^{-1}$ for long time. However, for the quench from the XX line to the incommensurate phase, the $\delta C_{mn}(t)$ is $\sim at^{-3/2}+bt^{-1}+ct^{1/2}$, which dominates for the scaling behavior of $\sim t^{-1/2}$. Both of them agree with the numerical results in Fig.~\ref{fig:delta.C.XX}. According to the relaxation behavior of $\delta C_{mn}(t)$, we plot the dynamical phase diagram as seen in Fig.~\ref{fig:dynamical.phase.diagram.XX}~(c). The boundary between the dynamical phases is the disorder line of the commensurate and incommensurate phases.

\section{Conclusion}

In this paper, we investigate the dynamical relaxation behavior of extended XY chains with the gapless phase after a quantum quench, in which the gapless phase is induced by the additional interactions: the DM interaction, XZY-YZX type of three-site interactions etc. This facilitates us to obtain the expression of the two-point correlation function $C_{mn}(t)$ in the quench from various initial states. We notice that in both models, the excitation spectrum satisfies  $\varepsilon_{k}+\varepsilon_{-k}=2\omega_{k}$, where $\omega_{k}$ is the excitation spectrum of the XY chain without additional interaction. This results in that when the quench is from the gapped phase, the additional interactions do not affect the relaxation behavior. The relaxation behavior is $\delta C_{mn}(t)\sim t^{-3/2}$ for the quench to the commensurate phase, and $\delta C_{mn}(t)\sim t^{-1/2}$ for the quench to the incommensurate phase.

In the case of the quench from the gapless phase, the initial state contains the single-occupied quasiparticle states, i.e. $|1_{k}0_{-k}\rangle$, which do not contribute to $\delta C_{mn}(t)$. This indicates that the additional interactions will affect the integral region of $\delta C_{mn}(t)$, and generate the boundary points in the asymptotic behavior. Consequently, we find the dynamical universal decay of the two-point correlation follows a power law of $t^{-1}$ and $t^{-1/2}$, where $t^{-1}$ is contributed by the integral around the boundary point. Specifically, when the quench is from the gapless phase to the commensurate phase, the power-law behavior of $\delta C_{mn}(t)$ is $t^{-1}$. However, when the quench is from the gapless phase to the incommensurate phase, there are two different cases. The one is that the interval $[k_{l}, k_{r}]$, in which $\varepsilon_{k}$ of the pre-quench Hamiltonian is smaller than zero, covering the minimum value of $\tilde{\omega}_{k}$ of the post-quench Hamiltonian. In this case, the power-law behavior is $t^{-1}$. The other is that the interval $[k_{l}, k_{r}]$ does not cover the $\min{\tilde{\omega}_{k}}$, where the power-law behavior is $t^{-1/2}$. Finally, we give the dynamical phase diagram and find it also depending on the position of pre-quench Hamiltonian.

In addition, we also study the case of quench from the XX line, in which the ground state contains the double-occupied quasiparticle states $|1_{k}1_{-k}\rangle$, due to the excitation spectrum satisfying the inverse symmetry with respect to $k=\pi$. The dynamical relaxation behavior of $\delta C_{mn}(t)$ is found to be $\sim t^{-1}$ for the quench from the XX line to the commensurate phase, and $\sim t^{-1/2}$ for the quench from the XX line to the incommensurate phase.

\begin{acknowledgments}
  The work is supported by the National Key Basic Research Program of China (No.~2020YFB0204800), the National Science Foundation of China (Grant Nos.~12204432, 11975126, and 12247106), and Key Research Projects of Zhejiang Lab (Nos. 2021PB0AC01 and 2021PB0AC02).
\end{acknowledgments}

\appendix
\begin{widetext}
\section{Correlation functions}

To obtain $C_{mn}(t)$, we transform the fermionic operators into the momentum space. We obtain
\begin{equation}
    \begin{split}
        C_{mn}(t) & = \langle\psi(t)| \frac{1}{N}\sum_{k}c_{k}^{\dag}c_{k}e^{ik(n-m)} |\psi(t)\rangle \\
        & = \frac{1}{N}\sum_{k>0}[\langle\psi_{k}(t)|c_{k}^{\dag}c_{k}|\psi_{k}(t)\rangle e^{ik(n-m)} + \langle\psi_{k}(t)|c_{-k}^{\dag}c_{-k}|\psi_{k}(t)\rangle e^{-ik(n-m)}],
    \end{split}
\end{equation}
where
    \begin{equation}\label{expectation.ckck}
        \begin{split}
            \langle\psi_{k}(t)|c_{k}^{\dag}c_{k}|\psi_{k}(t)\rangle & = \langle\psi_{k}(t)| (\cos{\tilde{\theta}_{k}}\tilde{\eta}_{k}^{\dag}+i\sin{\tilde{\theta}_{k}}\tilde{\eta}_{-k})(\cos{\tilde{\theta}_{k}}\tilde{\eta}_{k}-i\sin{\tilde{\theta}_{k}}\tilde{\eta}_{-k}^{\dag}) |\psi_{k}(t)\rangle \\
            & = \langle\psi_{k}(t)| \cos^{2}{\tilde{\theta}_{k}}\tilde{\eta}_{k}^{\dag}\tilde{\eta}_{k} -i\sin{\tilde{\theta}_{k}}\cos{\tilde{\theta}_{k}}\tilde{\eta}_{k}^{\dag}\tilde{\eta}_{-k}^{\dag} + i\sin{\tilde{\theta}_{k}}\cos{\tilde{\theta}_{k}}\tilde{\eta}_{-k}\tilde{\eta}_{k} + \sin^{2}{\tilde{\theta}_{k}}\tilde{\eta}_{-k}\tilde{\eta}_{-k}^{\dag} |\psi_{k}(t)\rangle
        \end{split}
    \end{equation}
    and
    \begin{equation}\label{expectation.c-kc-k}
        \begin{split}
            \langle\psi_{k}(t)|c_{-k}^{\dag}c_{-k}|\psi_{k}(t)\rangle & = \langle\psi_{k}(t)| (\cos{\tilde{\theta}_{k}}\tilde{\eta}_{-k}^{\dag}-i\sin{\tilde{\theta}_{k}}\tilde{\eta}_{k})(\cos{\tilde{\theta}_{k}}\tilde{\eta}_{-k}+ i\sin{\tilde{\theta}_{k}}\tilde{\eta}_{k}^{\dag}) |\psi_{k}(t)\rangle \\
            & = \langle\psi_{k}(t)| \cos^{2}{\tilde{\theta}_{k}}\tilde{\eta}_{-k}^{\dag}\tilde{\eta}_{-k} + i\sin{\tilde{\theta}_{k}}\cos{\tilde{\theta}_{k}}\tilde{\eta}_{-k}^{\dag}\tilde{\eta}_{k}^{\dag} - i\sin{\tilde{\theta}_{k}}\cos{\tilde{\theta}_{k}}\tilde{\eta}_{k}\tilde{\eta}_{-k} + \sin^{2}{\tilde{\theta}_{k}}\tilde{\eta}_{k}\tilde{\eta}_{k}^{\dag} |\psi_{k}(t)\rangle.
        \end{split}
    \end{equation}

To calculate $C_{mn}(t)$, we need to calculate every component in Eqs. (\ref{expectation.ckck}) and (\ref{expectation.c-kc-k}). According to Eq.~(\ref{time-evolved.state.k}), we can obtain
\begin{equation}\label{component.kk}
    \langle\psi_{k}(t)|\tilde{\eta}_{k}^{\dag}\tilde{\eta}_{k}|\psi_{k}(t)\rangle = 1-\langle\psi_{k}(t)|\tilde{\eta}_{k}\tilde{\eta}_{k}^{\dag}|\psi_{k}(t)\rangle = \left\{
      \begin{array}{l}
        \sin^{2}{\alpha_{k}}, \quad \varepsilon_{k},\varepsilon_{-k}>0, \\
        0, \quad \varepsilon_{k}>0, \varepsilon_{-k}\leq0, \\
        1, \quad \varepsilon_{k}\leq0,\varepsilon_{-k}>0, \\
        \cos^{2}{\alpha_{k}}, \quad \varepsilon_{k},\varepsilon_{-k}\leq0,
      \end{array}
    \right.
\end{equation}
\begin{equation}\label{component.k-kdag}
    \langle\psi_{k}(t)|\tilde{\eta}_{k}^{\dag}\tilde{\eta}_{-k}^{\dag}|\psi_{k}(t)\rangle = -\langle\psi_{k}(t)|\tilde{\eta}_{-k}^{\dag}\tilde{\eta}_{k}^{\dag}|\psi_{k}(t)\rangle = \left\{
      \begin{array}{l}
        i\sin{\alpha_{k}}\cos{\alpha_{k}}e^{it(\tilde{\varepsilon}_{k}+\tilde{\varepsilon}_{-k})}, \varepsilon_{k},\varepsilon_{-k}>0, \\
        0, \quad \varepsilon_{k}>0, \varepsilon_{-k}\leq0, \\
        0, \quad \varepsilon_{k}\leq0,\varepsilon_{-k}>0, \\
        -i\sin{\alpha_{k}}\cos{\alpha_{k}}e^{it(\tilde{\varepsilon}_{k}+\tilde{\varepsilon}_{-k})}, \quad \varepsilon_{k},\varepsilon_{-k}\leq0,
      \end{array}
    \right.
\end{equation}
\begin{equation}\label{component.k-k}
    \langle\psi_{k}(t)| \tilde{\eta}_{-k}\tilde{\eta}_{k} |\psi_{k}(t)\rangle = -\langle\psi_{k}(t)| \tilde{\eta}_{k}\tilde{\eta}_{-k} |\psi_{k}(t)\rangle = \left\{
      \begin{array}{l}
        -i\sin{\alpha_{k}}\cos{\alpha_{k}}e^{-it(\tilde{\varepsilon}_{k}+\tilde{\varepsilon}_{-k})}, \varepsilon_{k},\varepsilon_{-k}>0, \\
        0, \quad \varepsilon_{k}>0, \varepsilon_{-k}\leq0, \\
        0, \quad \varepsilon_{k}\leq0,\varepsilon_{-k}>0, \\
        i\sin{\alpha_{k}}\cos{\alpha_{k}}e^{-it(\tilde{\varepsilon}_{k}+\tilde{\varepsilon}_{-k})}, \quad \varepsilon_{k},\varepsilon_{-k}\leq0,
      \end{array}
    \right.
\end{equation}
\begin{equation}\label{component.-k-k}
    \langle\psi_{k}(t)| \tilde{\eta}_{-k}\tilde{\eta}_{-k}^{\dag} |\psi_{k}(t)\rangle = 1-\langle\psi_{k}(t)| \tilde{\eta}_{-k}^{\dag}\tilde{\eta}_{-k} |\psi_{k}(t)\rangle = \left\{
      \begin{array}{l}
        \cos^{2}{\alpha_{k}}, \varepsilon_{k},\varepsilon_{-k}>0, \\
        0, \quad \varepsilon_{k}>0, \varepsilon_{-k}\leq0, \\
        1, \quad \varepsilon_{k}\leq0,\varepsilon_{-k}>0, \\
        \sin^{2}{\alpha_{k}}, \quad \varepsilon_{k},\varepsilon_{-k}\leq0.
      \end{array}
    \right.
\end{equation}
Substituting Eqs.~(\ref{component.kk}, \ref{component.k-kdag}, \ref{component.k-k}, \ref{component.-k-k}) into Eqs.~(\ref{expectation.ckck}, \ref{expectation.c-kc-k}), we will have
\begin{equation}
    \langle\psi_{k}(t)|c_{k}^{\dag}c_{k}|\psi_{k}(t)\rangle = \left\{
      \begin{array}{l}
        \cos^{2}{\tilde{\theta}_{k}}\sin^{2}{\alpha_{k}} + \frac{1}{2}\sin{2\tilde{\theta}_{k}}\sin{2\alpha_{k}}\cos{[(\tilde{\varepsilon}_{k}+\tilde{\varepsilon}_{-k})t]}+\sin^{2}{\tilde{\theta}_{k}}\cos^{2}{\alpha_{k}}, \quad \varepsilon_{k},\varepsilon_{-k}>0, \\
        0, \quad \varepsilon_{k}>0, \varepsilon_{-k}\leq0, \\
        1, \quad \varepsilon_{k}\leq0,\varepsilon_{-k}>0, \\
        \cos^{2}{\tilde{\theta}_{k}}\cos^{2}{\alpha_{k}} - \frac{1}{2}\sin{2\tilde{\theta}_{k}}\sin{2\alpha_{k}}\cos{[(\tilde{\varepsilon}_{k}+\tilde{\varepsilon}_{-k})t]}+\sin^{2}{\tilde{\theta}_{k}}\sin^{2}{\alpha_{k}} , \quad \varepsilon_{k},\varepsilon_{-k}\leq0,
      \end{array}
    \right.
\end{equation}
and
\begin{equation}
    \langle\psi_{k}(t)|c_{-k}^{\dag}c_{-k}|\psi_{k}(t)\rangle = \left\{
      \begin{array}{l}
        \cos^{2}{\tilde{\theta}_{k}}\sin^{2}{\alpha_{k}} + \frac{1}{2}\sin{2\tilde{\theta}_{k}}\sin{2\alpha_{k}}\cos{[(\tilde{\varepsilon}_{k}+\tilde{\varepsilon}_{-k})t]}+\sin^{2}{\tilde{\theta}_{k}}\cos^{2}{\alpha_{k}}, \varepsilon_{k},\varepsilon_{-k}>0, \\
        0, \quad \varepsilon_{k}>0, \varepsilon_{-k}\leq0, \\
        1, \quad \varepsilon_{k}\leq0,\varepsilon_{-k}>0, \\
        \cos^{2}{\tilde{\theta}_{k}}\cos^{2}{\alpha_{k}} - \frac{1}{2}\sin{2\tilde{\theta}_{k}}\sin{2\alpha_{k}}\cos{[(\tilde{\varepsilon}_{k}+\tilde{\varepsilon}_{-k})t]}+\sin^{2}{\tilde{\theta}_{k}}\sin^{2}{\alpha_{k}} , \quad \varepsilon_{k},\varepsilon_{-k}\leq0.
      \end{array}
    \right.
\end{equation}
These indicate that correlation functions are independent of time for the single-occupied states. For $\varepsilon_{k}, \varepsilon_{-k} > 0$, we have
\begin{equation}
    \begin{split}
        C_{mn}^{k}(t) & =
        \langle\psi_{k}(t)|c_{k}^{\dag}c_{k}|\psi_{k}(t)\rangle e^{ik(n-m)}+\langle\psi_{k}(t)|c_{-k}^{\dag}c_{-k}|\psi_{k}(t)\rangle e^{-ik(n-m)} \\
        & = [\langle\psi_{k}(t)|c_{k}^{\dag}c_{k}|\psi_{k}(t)\rangle + \langle\psi_{k}(t)|c_{-k}^{\dag}c_{-k}|\psi_{k}(t)\rangle]\cos{[k(n-m)]} \\
        & + i [\langle\psi_{k}(t)|c_{k}^{\dag}c_{k}|\psi_{k}(t)\rangle - \langle\psi_{k}(t)|c_{-k}^{\dag}c_{-k}|\psi_{k}(t)\rangle]\sin{[k(n-m)]} \\
        & = \{1 - \cos{2\tilde{\theta}_{k}}\cos{2\alpha_{k}} + \sin{2\tilde{\theta}_{k}}\sin{2\alpha_{k}}\cos{[(\tilde{\varepsilon}_{k}+\tilde{\varepsilon}_{-k})t]}\}\cos{[k(n-m)]},
    \end{split}
\end{equation}
and for $\varepsilon_{k}, \varepsilon_{-k}\leq0$,
\begin{equation}
    C_{mn}^{k}(t) = \{1 + \cos{2\tilde{\theta}_{k}}\cos{2\alpha_{k}} - \sin{2\tilde{\theta}_{k}}\sin{2\alpha_{k}}\cos{[(\tilde{\varepsilon}_{k}+\tilde{\varepsilon}_{-k})t]}\}\cos{[k(n-m)]}.
\end{equation}
$C_{mn}^{k}(t)$ consists of two components: one is the value of $C_{mn}^{k}(t)$ in the steady state, i.e.
\begin{equation}\label{steady.state.value}
    C_{mn}^{k}(\infty) = \left\{
      \begin{array}{l}
           (1-\cos{2\tilde{\theta}_{k}}\cos{2\alpha_{k}})\cos{[k(n-m)]}, \quad \varepsilon_{k},\varepsilon_{-k}>0,  \\
           (1+\cos{2\tilde{\theta}_{k}}\cos{2\alpha_{k}})\cos{[k(n-m)]}, \quad \varepsilon_{k},\varepsilon_{-k}\leq0;
      \end{array}
    \right.
\end{equation}
and the other one is the difference between $C_{mn}^{k}(t)$ and $C_{mn}^{k}(\infty)$, i.e.
\begin{equation}\label{delta.t.value}
    \delta C_{mn}^{k}(t) = \left\{
      \begin{array}{l}
           \sin{2\tilde{\theta}_{k}}\sin{2\alpha_{k}}\cos{[(\tilde{\varepsilon}_{k}+\tilde{\varepsilon}_{-k})t]}\cos{[k(n-m)]}, \quad \varepsilon_{k},\varepsilon_{-k}>0,  \\
           - \sin{2\tilde{\theta}_{k}}\sin{2\alpha_{k}}\cos{[(\tilde{\varepsilon}_{k}+\tilde{\varepsilon}_{-k})t]}\cos{[k(n-m)]}, \quad \varepsilon_{k},\varepsilon_{-k}\leq0.
      \end{array}
    \right.
\end{equation}

\end{widetext}

\section{Stationary point approximation for the case from the gapped phase}

In the following, we use the stationary phase approximation to explain the relaxation behavior.

For quenching from the gapped phase to the commensurate phase, there are two stationary points $k_{0}=0, \pi$. By considering the Bogoliubov angles satisfy $\tan{2\theta_{k}}=\frac{\gamma\sin{k}}{h+\cos{k}}$, we have
\begin{equation}\label{eq:saddle-f}
    f(k) = \frac{[h_{1}\gamma_{0}-h_{0}\gamma_{1}+(\gamma_{0}-\gamma_{1})\cos{k}]\gamma_{1}\sin^{2}{k}}{\omega_{k}\tilde{\omega}_{k}^{2}},
\end{equation}
so that $f(0)$, $f'(0)$, $f(\pi)$ and $f'(\pi)$ vanish. We thus need to expand $f(k)$ around $k_{0}$ and go to the second-order contribution $(\gamma_{0}=\gamma_{1})$
\begin{equation}\label{eq:f-gapped-commen}
    \begin{split}
        f(k) = \frac{-(h_{1}-h_{0})\gamma_{1}^{2}}{\omega_{k_{0}}\tilde{\omega}_{k_{0}}^{2}}(k-k_{0})^{2}.
    \end{split}
\end{equation}
Considering $\omega_{0}(\tilde{\omega}_{0})>\omega_{\pi}(\tilde{\omega}_{\pi})$, the contribution of the stationary point $k_{0}=0$ is quite smaller than that of $k_{0}=\pi$. Hence, the approximate behavior of $I(t)$ is determined by the contribution of the stationary point $k_{0}=\pi$, i.e.
\begin{equation}
    \begin{split}
        I(t) & \approx \frac{1}{2\pi}\frac{-(h_{1}-h_{0})\gamma_{1}^{2}}{\omega_{\pi}\tilde{\omega}_{\pi}^{2}}e^{2it\tilde{\omega}_{\pi}} \\
             & \quad\quad\quad\quad\quad\quad\quad \cdot \int_{0}^{+\infty} dk (k-\pi)^{2}e^{it\tilde{\omega}^{''}_{\pi}(k-\pi)^{2}} \\
             & = \frac{1}{2\pi}\frac{-(h_{1}-h_{0})\gamma_{1}^{2}}{\omega_{\pi}\tilde{\omega}_{\pi}^{2}}e^{2it\tilde{\omega}_{\pi}+i\phi}\sqrt{\frac{\pi}{\tilde{\omega}_{\pi}^{''}}}t^{-3/2}.
    \end{split}
\end{equation}
The asymptotic behavior of $\delta C_{mn}(t)$ is thus given by
\begin{equation}
  \delta C_{mn}(t) = \frac{-(h_{1}-h_{0})\gamma_{1}^{2}}{2\omega_{\pi}\tilde{\omega}_{\pi}^{2}\sqrt{\pi(\tilde{\omega}_{\pi}^{''})^{3}}}\cos{(2\tilde{\omega}_{\pi}t+\phi)}t^{-3/2}.
\end{equation}
Here, the cosine terms describes the oscillation of $\delta C_{mn}(t)$, so that the decay of $\delta C_{mn}(t)$ is given by $\sim t^{-3/2}$, which agrees with the numerical simulations in Fig.~\ref{fig:delta_C_gapped}~(a).

While for quenching from the gapped phase to the incommensurate phase, there is an additional stationary point of $\tilde{\omega}_{k_{m}}$ for $0<k_{m}<\pi$ besides $k_{0}=0,\pi$, where $\tilde{\omega}_{k_{m}}$ is the minimum value of $\tilde{\omega}_{k}$. $I(t)$ should be calculated by summing the integrals over all stationary points, i.e.
\begin{equation}
  I(t) = I_{1}(t)+ I_{2}(t)+ I_{3}(t),
\end{equation}
where $I_{1}(t), I_{2}(t)\sim t^{-3/2}$ denotes the integrals around the stationary points $k_{0}=0,\pi$, and
\begin{equation}\label{eq:I3_gapped}
    \begin{split}
      I_{3}(t) & \approx \frac{(h_{1}-h_{0})\gamma_{1}^{2}\sin^{2}{k_{m}}}{2\pi\omega_{k_{m}}\tilde{\omega}_{k_{m}}^{2}}e^{2i\tilde{\omega}_{k_{m}}}\int_{m}^{+\infty}dk  e^{it\tilde{\omega}_{k_{m}}^{''}(k-k_{m})^{2}} \\
               & = \frac{(h_{1}-h_{0})\gamma_{1}^{2}\sin^{2}{k}}{2\omega_{k_{m}}\tilde{\omega}_{k_{m}}^{2}\sqrt{\pi\tilde{\omega}_{k_{m}}^{''}}}e^{2it\tilde{\omega}_{k_{m}}+i\phi}t^{-1/2} \\
               & \sim t^{-1/2}
    \end{split}
\end{equation}
denotes the integral around the minimum value of $\tilde{\omega}_{k}$.

Considering $t^{-3/2}$ decays faster than $t^{-1/2}$, the approximate behavior of $\delta C_{mn}(t)$ is determined by the slowest decay, i.e. $\delta C_{mn}(t)\sim t^{-1/2}$. This result also agrees with the numerical simulations in Fig.~\ref{fig:delta_C_gapped}~(b).

\section{Boundary of the dynamical phase diagram}

The second condition to distinguish the dynamical phase diagram is whether the interval $[k_{l}, k_{r}]$ covers the minimum value of $\tilde{\omega}_{k}$, which can be expressed by the following equations
\begin{equation}\label{eq:diagram.condition}
  \left\{
  \begin{array}{l}
    \varepsilon_{k} = 0, \\
    \cos{k}=\frac{h}{\gamma^{2}-1}.
  \end{array}
  \right.
\end{equation}
Here $\varepsilon_{k}$ is the excitation spectrum of the pre-quench Hamiltonian $H(h_{0}, \gamma_{0})$.

For the XY chain with DM interaction, the Eq.~(\ref{eq:diagram.condition}) reduces to
\begin{equation}
  \left\{
    \begin{array}{l}
      -2D\sin{k}+\sqrt{(h_{0}+\cos{k})^{2}+\gamma_{0}^{2}\sin^{2}{k}}=0, \\
      \cos{k} = -\frac{h}{1-\gamma^{2}},
    \end{array}
  \right.
\end{equation}
which can be written as a quadratic equation of $\frac{h}{1-\gamma^{2}}$ by
\begin{equation}\label{eq:quadratic.equation.dm}
  (1-\gamma_{0}^{2}+4D^{2})(\frac{h}{1-\gamma^{2}})^{2}-2h_{0}\frac{h}{1-\gamma^{2}}+h_{0}^{2}+\gamma_{0}^{2}-4D^{2}=0
\end{equation}
The Eq.~(\ref{eq:quadratic.equation.dm}) can be solved by
\begin{equation}
  \frac{h}{1-\gamma^{2}} = \frac{h_{0}\pm\sqrt{h_{0}^{2}-(1-\gamma_{0}^{2}+4D^{2})(h_{0}^{2}+\gamma_{0}^{2}-4D^{2})}}{1-\gamma_{0}^{2}+4D^{2}}.
\end{equation}
Therefore, the boundary of dynamical phase in the case of quench from the gapless phase is also dependent of the position of the pre-quench Hamiltonian. Similar conclusion can also be obtained for the XY chain with the XZY-YZX type of three-site interaction.


\bibliography{non-equilibrium}

\end{document}